\documentclass[12pt,english]{article}

\usepackage{xargs}                      
\usepackage[pdftex,dvipsnames]{xcolor}  
\usepackage[colorinlistoftodos,prependcaption,textsize=tiny]{todonotes}

\newcommandx{\unsure}[2][1=]{\todo[linecolor=red,backgroundcolor=red!25,bordercolor=red,#1]{#2}}
\newcommandx{\change}[2][1=]{\todo[linecolor=blue,backgroundcolor=blue!25,bordercolor=blue,#1]{#2}}
\newcommandx{\info}[2][1=]{\todo[linecolor=OliveGreen,backgroundcolor=OliveGreen!25,bordercolor=OliveGreen,#1]{#2}}
\newcommandx{\improvement}[2][1=]{\todo[linecolor=Plum,backgroundcolor=Plum!25,bordercolor=Plum,#1]{#2}}
\newcommandx{\thiswillnotshow}[2][1=]{\todo[disable,#1]{#2}}

\usepackage{amssymb,amsmath}

\usepackage[verbose,letterpaper,tmargin=2.54cm,bmargin=2.54cm,lmargin=2.54cm,rmargin=2.54cm]{geometry}
\usepackage{setspace}
\usepackage[displaymath,mathlines]{lineno}

\usepackage[affil-it]{authblk}
\providecommand{\keywords}[1]{\indent \textit{Key words: #1}}

\usepackage{lmodern}

\usepackage{ifxetex,ifluatex}
\usepackage{fixltx2e} 
\ifnum 0\ifxetex 1\fi\ifluatex 1\fi=0 
  \usepackage[T1]{fontenc}
  \usepackage[utf8]{inputenc}
\else 
  \ifxetex
    \usepackage{mathspec}
    \usepackage{xltxtra,xunicode}
  \else
    \usepackage{fontspec}
  \fi
  \defaultfontfeatures{Mapping=tex-text,Scale=MatchLowercase}
  
\fi
\IfFileExists{upquote.sty}{\usepackage{upquote}}{}
\IfFileExists{microtype.sty}{\usepackage{microtype}}{}
\usepackage{natbib}

\usepackage{graphicx}
\makeatletter
\def\maxwidth{\ifdim\Gin@nat@width>\linewidth\linewidth\else\Gin@nat@width\fi}
\def\maxheight{\ifdim\Gin@nat@height>\textheight\textheight\else\Gin@nat@height\fi}
\makeatother
\setkeys{Gin}{width=\maxwidth,height=\maxheight,keepaspectratio}
\ifxetex
  \usepackage[setpagesize=false, 
              unicode=false, 
              xetex]{hyperref}
\else
  \usepackage[unicode=true]{hyperref}
\fi
\hypersetup{breaklinks=true,
            bookmarks=true,
            pdfauthor={},
            pdftitle={Generalizing the first-difference correlated random walk for marine animal movement data},
            colorlinks=true,
            citecolor=blue,
            urlcolor=blue,
            linkcolor=magenta,
            pdfborder={0 0 0}}
\title{Generalizing the first-difference correlated random walk for marine
animal movement data}

\author[]{Christoffer Moesgaard Albertsen}

\affil[]{National Institute of Aquatic Resources, Technical University of
Denmark, Kemitorvet 201, DK-2800 Kgs. Lyngby, Denmark}

\date{cmoe@aqua.dtu.dk}
\usepackage{longtable}
\usepackage{booktabs}
\usepackage{graphicx}
\usepackage{bbm}
\usepackage{colortbl}

\begin{document}
\begin{spacing}{2.0}
\maketitle
\begin{abstract}
Animal telemetry data are often analysed with discrete time movement
models assuming rotation in the movement. These models are defined with
equidistant distant time steps. However, telemetry data from marine
animals are observed irregularly. To account for irregular data, a
time-irregularised first-difference correlated random walk model with
drift is introduced. The model generalizes the commonly used
first-difference correlated random walk with regular time steps by
allowing irregular time steps, including a drift term, and by allowing
different autocorrelation in the two coordinates. The model is applied
to data from a ringed seal collected through the Argos satellite system,
and is compared to related movement models through simulations.
Accounting for irregular data in the movement model results in accurate
parameter estimates and reconstruction of movement paths. Measured by
distance, the introduced model can provide more accurate movement paths
than the regular time counterpart. Extracting accurate movement paths
from uncertain telemetry data is important for evaluating space use
patterns for marine animals, which in turn is crucial for management.
Further, handling irregular data directly in the movement model allows
efficient simultaneous analysis of several animals.

\keywords{animal telemetry, movement ecology, correlated random walk, irregular
data}
\end{abstract}

\hypertarget{introduction}{%
\section{Introduction}\label{introduction}}

Understanding animal movement behaviour, space use patterns and response
to the environment relies on modelling animal telemetry data. So far,
the collection of telemetry data has been hindered in marine
environments. Satellite systems require animals to surface, storage tags
require recapture, and acoustic networks require a dense array of
receivers. Nonetheless, the scope and scales of movement studies are
continuously increasing with rapid technological advances in tracking
devices \citep{hussey2015a}.

In spite of technological advances leading to more accurate measurements
and larger data sets, data from satellite tags have inherent measurement
errors. Systems such as Fastloc GPS and Argos can only record data when
the animal is above water, with an accuracy that depends on the surface
time and diving behaviour \citep{costa2010a}. Typically, this results in
data observed at irregular time steps with considerable uncertainty,
which makes state-space models a valuable framework for analysing the
data.

State-space models naturally separate the movement process from
measurement errors. True animal locations are explicitly modelled
separately from the observed measurements conditionally on the
locations. Hence, the movement models developed can be combined with
telemetry data from any source, for example, GPS, Argos, light based, or
acoustic tags. However, inference in general state-space models can be
challenging. Maximum likelihood estimation in state-space models
requires integration of the joint distribution of locations and
measurements over the locations to obtain the marginal distribution of
the observations. Recent software such as AD Model Builder
\citep{fournier2012a} and Template Model Builder
\citep[\texttt{TMB};][]{kristensen2016a} utilizes the Laplace
approximation to rapidly approximate the marginal distribution of
observations in state-space models. While care is needed to ensure the
approximation is appropriate, this framework is generally applicable.
For instance, \texttt{TMB} has been used for both discrete time
\citep{augermethe2016a, augermethe2017a} and continuous time
\citep{albertsen2015a} movement models. Alternatively, the state-space
can be discretized and the resulting hidden Markov model can be used for
inference \citep{pedersen2011a}.

Although continuous time movement models can handle irregular data,
discrete time models are often thought of as more intuitive
\citep{mcclintock2014a}. Further, continuous time models can be
difficult to construct or extend analytically, because they are often
formulated through stochastic differential equations. Whereas continuous
time models describe instantaneous change, discrete time models describe
the change between two observations. In animal movement models, the
change between observations are often modelled through a step length and
an angle of movement
\citep[e.g.,][]{morales2004a, jonsen2005a, guarie2009a, tracey2010a, mcclintock2012a, michelot2016a},
which is considered a natural way to represent movement
\citep{turchin1998a}. In discrete time models with regular time steps,
irregular data can be handled by modifying the observational model to
interpolate between states \citep{jonsen2005a}.

To facilitate the analysis of populations or meta-analysis of
individuals, the model parameters for different individuals must be on
the same scale. If the interpretation of parameter values depends on the
chosen time steps between estimated locations \citep{mcclintock2014a},
the values must be corrected before comparison is possible. Likewise,
estimating population parameters based on individuals either requires
movement models that use the same time steps for all individuals
\citep[e.g.][]{jonsen2016a}, which can be sub-optimal, or movement
models that handle irregular data directly.

The first-difference correlated random walk \citep[DCRW;][]{jonsen2005a}
models the animal movement as a discrete time first order
auto-regressive process on the difference between consecutive locations
through a rotation matrix. A continuous time version of the DCRW without
rotation is the continuous time correlated random walk
\citep[CTCRW;][]{johnson2008a}. The CTCRW models the velocity of an
animal by an Ornstein-Uhlenbeck process: the continuous time counterpart
of the first order auto-regressive process. Recently, an irregular time
version of the DCRW was introduced without rotation including a
time-varying parameter instead \citep{augermethe2017a}.

In this paper, a generalization of the DCRW is introduced. The
generalization of the DCRW is trifold: the model allows irregular time
steps, a drift vector is included, and different auto-correlations in
the two coordinates can be used. As a by-product, it is shown how
parameters of the DCRW model depends on the selected time steps, and how
to transform them to a common scale. The generalized DCRW is introduced
as the discretization of a stochastic differential equation for the
animal's velocity and location. Further, the close relation to a
generalization of the CTCRW is shown. Through three simulation studies,
the generalized DCRW is shown to perform well compared to the CTCRW and
DCRW. Finally, The applicability and extendability of the model is
illustrated through a real data set collected by the Argos system.

\hypertarget{materials-and-methods}{%
\section{Materials and Methods}\label{materials-and-methods}}

\hypertarget{movement-model}{%
\subsection{Movement model}\label{movement-model}}

The generalization of the DCRW movement model is constructed through a
stochastic differential equation (SDE) for the velocity. From this SDE,
the location process is the integrated velocity process. To obtain the
time-irregular movement model, the SDE is solved analytically while the
location process is a discrete time approximation.

\hypertarget{sde-for-velocity}{%
\subsubsection{SDE for velocity}\label{sde-for-velocity}}

The bivariate SDE for the velocity, \(V_t\), of an animal is
\begin{equation}
\begin{aligned}
\text{d}V_t &= -\left(\begin{array}{cc} -\log\gamma_1 & \theta \\ -\theta & -\log\gamma_2 \end{array}\right) (V_t-\mu) \text{d}t + S dB_t \\
&= - \Theta (V_t-\mu) \text{d}t + S dB_t
\end{aligned}
\label{eq:sdeV}\end{equation} with initial condition
\(V_0 = v_0 \in \mathbb{R}^2\). In this model, \(\gamma_1\) and
\(\gamma_2\) are autocorrelation parameters, \(\mu\) is a vector of mean
velocity parameters, and \(\theta\) is the mean turning angle of the
movement process. The matrix \(S\) is a lower triangular matrix with
positive diagonal determining the covariance of the changes in velocity.
Animals moving persistently in one direction will have high
autocorrelation in the velocity and a mean turning angle close to zero,
while animals foraging or in other ways displaying a tortuous movement
will have a lower autocorrelation in the velocity and a mean turning
angle different from zero (modulo \(2\pi\))
\citep[e.g.][]{jonsen2005a, whoriskey2017a}.

The analytical solution to the SDE is a stochastic process with Gaussian
increments (see e.g.~Supplemental Material). The mean of an increment is
\[
E(V_{t} \mid V_{s}) = \exp(-\Theta (t-s)) V_s + \left(I-e^{-\Theta (t-s)}\right)\mu,\quad s<t
\] where \(\mu = (\mu_1,\mu_2)^T\), while the covariance is \[
\operatorname{vec}(Var\left( V_{t} \mid V_{s} \right)) = \operatorname{vec}(C) - e^{-\Theta (t-s)}\otimes e^{-\Theta (t-s)} \operatorname{vec}(C),\quad s<t
\] where
\(\operatorname{vec}(C) = (\Theta\oplus\Theta)^{-1} \operatorname{vec}(\Sigma)\),
\(\Sigma = SS^T\), \(\oplus\) denotes the Kronecker sum,
\(A \oplus B = A \otimes I_B + I_A \otimes B\), and
\(\operatorname{vec}\) is an operator that stacks the columns of a
matrix to a vector. Based on the velocity, the location process,
\(X_t\), follows the SDE \begin{equation}
\text{d}X_t = V_t\text{d}t 
\label{eq:sdeX}\end{equation}

\hypertarget{discretizing-the-sde-for-locations}{%
\subsubsection{Discretizing the SDE for
locations}\label{discretizing-the-sde-for-locations}}

Considering the \(N\) predetermined time points
\(\{t_i\}_{i\in \{1,\ldots,N\}}\), an Euler--Maruyama approximation to
the location process is obtained by \[
X_{t_i} = X_{t_{i-1}} + V_{t_{i-1}} \Delta_{i}
\] where \(\Delta_{i} = t_i - t_{i-1}\) is the length of the time step.
Inserting the expression for \(V_{t_{i-1}}\), the Generalized
first-Differenced Correlated Random Walk (GDCRW) is obtained:
\begin{equation}
X_{t_i} = X_{t_{i-1}} + \Delta_{i} \exp(-\Theta \Delta_{i-1}) (X_{t_{i-1}} - X_{t_{i-2}})/\Delta_{{i-1}} + \Delta_{i}\left(I-\exp(-\Theta \Delta_{i-1})\right)\mu + \Delta_{i} \epsilon_{t_i}
\label{eq:gdcrw}\end{equation} The error terms, \(\epsilon_{t_i}\),
follow a zero mean normal distribution with variance \[
Var(\epsilon_{t_i}) = C - \exp(-\Theta \Delta_{i}) C \exp(-\Theta^T \Delta_{i})
\] where \[
\operatorname{vec}(C) = (\Theta \oplus \Theta) ^ {-1} \operatorname{vec}(\Sigma),
\] and \(\Sigma = SS^T\) is a covariance matrix.

\hypertarget{special-cases}{%
\subsubsection{Special cases}\label{special-cases}}

The GDCRW model is closely related to four other models: The CTCRW, the
DCRW, the modified DCRW by \citet{augermethe2017a}, and a random walk.

\hypertarget{relation-to-the-ctcrw}{%
\paragraph{\texorpdfstring{Relation to the CTCRW
\label{gdcrwVSctcrw}}{Relation to the CTCRW }}\label{relation-to-the-ctcrw}}

When the turning-angle parameter \(\theta = 0\), and the increment
covariance \(S\) is diagonal, the velocity model, \(V_t\) is identical
to the velocity model of the CTCRW; however, the location models differ.
While the location model of the CTCRW is an analytical solution to the
two univariate integrated velocity models, the GDCRW is a discretization
of the more general bivariate velocity model. Hence, when the length of
the time steps approaches zero, the GDCRW approaches the CTCRW when the
velocity models are identical. When \(\theta \neq 0\) or \(S\) is not
diagonal, the GDCRW approaches a generalization of the CTCRW when the
length of the time steps approaches zero.

\hypertarget{relation-to-the-dcrw}{%
\paragraph{Relation to the DCRW}\label{relation-to-the-dcrw}}

For regularly observed data, \(\Delta_t = \Delta\), with \(\mu = 0\) and
\(\gamma_1 = \gamma_2\), \[
\exp(-\Theta \Delta) = \left(\begin{array}{cc} \cos(\theta\Delta) & -\sin(\theta\Delta) \\ \sin(\theta\Delta) & \cos(\theta\Delta) \end{array}\right) \gamma^\Delta
\] Hence, the movement model simplifies to
\begin{equation}X_i = X_{i-1} + \left(\begin{array}{cc} \cos(\theta\Delta) & -\sin(\theta\Delta) \\ \sin(\theta\Delta) & \cos(\theta\Delta) \end{array}\right) \gamma^\Delta (X_{i-1} - X_{i-2}) + \Delta\epsilon_i,\label{eq:reggdcrw}\end{equation}
which is the movement model of \citet{jonsen2005a} when \(\Delta = 1\).
Therefore, the GDCRW generalizes the DCRW in three ways: the restriction
of regular time steps is relaxed, a drift vector is added, and different
auto-correlations in the two coordinates are allowed. Further, equation
(\ref{eq:reggdcrw}) allows scaling the DCRW parameters to a common time
scale. That is, if parameters \(\tilde{\gamma}\) and \(\tilde{\theta}\)
are obtained from a DCRW model fitted with time steps
\(\tilde{\Delta}\), scale independent parameters are obtained by
\(\gamma = \tilde{\gamma}^{1/\tilde{\Delta}}\) and
\(\theta = \tilde{\theta}/\tilde{\Delta}\). Being able to scale the DCRW
parameters to a common scale allows comparison between model fits in
meta or population studies even if different time scales are used.

\hypertarget{relation-to-the-modified-dcrw}{%
\paragraph{Relation to the modified
DCRW}\label{relation-to-the-modified-dcrw}}

Recently, \citet{augermethe2017a} proposed a modified version of the
DCRW for Fastloc-GPS data. The modified DCRW allowed irregularly
observed data and included a time-varying autocorrelation parameter,
\(\gamma_t\). Ignoring the time-varying autocorrelation, the modified
DCRW can be obtained as a special case of the GDCRW introduced here by
letting \(\gamma_1 = \gamma_2 = \gamma_t\), \(\mu = 0\), \(\theta = 0\),
and by letting \(\Sigma\) be a diagonal matrix. Note, however, that the
two models differ slightly in parameterization. In the modified DCRW,
\(\gamma_t\) is not scaled by the time differences and the variance of
\(\epsilon_{t_i}\) is parameterized as a diagonal matrix
\(diag(\sigma_1^2, \sigma_2^2)\). The GDCRW can easily be extended to
have a time-varying autocorrelation parameter in the same way as
\citet{augermethe2017a}.

\hypertarget{relation-to-a-random-walk}{%
\paragraph{Relation to a random walk}\label{relation-to-a-random-walk}}

In the same manner as the three models above, the GDCRW can be reduced
to a random walk model. From equation (\ref{eq:gdcrw}) it follows that
when the autocorrelation parameters \(\gamma_1\) and \(\gamma_2\) tend
to zero, the model is reduced to a random walk. If \(\mu \neq 0\), it
will be a random walk with drift.

\hypertarget{measurement-equation}{%
\subsection{Measurement equation}\label{measurement-equation}}

Besides the movement process describing the latent states,
\(\{X_{t_i}\}_{i\in\{1,\ldots,N\}}\), a measurement equation is needed
to form a state-space model. For the DCRW, measurement errors are often
included by linearly interpolating between the latent states. For an
observation \(Y_{s_j}\) at time \(s_j\), \[
Y_{s_j} = (1 - q) X_{t_i}  + q X_{t_{i+1}} + \eta_j
\] such that \(t_i \leq s_j < t_{i+1}\),
\(q = \frac{s_j - t_i}{\Delta_{i+1}}\), and \(\eta_i\) is a zero mean
bivariate random variable. However, when the movement model allows
irregular time steps, it is more natural to ensure that the latent
states align with the observations, such that the measurement equation
simplifies to \[
Y_{s_j} = X_{s_j} + \eta_j
\] Then the observation at time \(s_j\) does not depend on a latent
state in the future. Note that the state-space model can include
additional latent states between observations. These latent states only
contributes to the likelihood function through the movement process.
Below, \(\eta_j\) will either be modelled as a bivariate normal
distribution \citep{johnson2008a} or a bivariate t-distribution
\citep{albertsen2015a}.

\hypertarget{simulation-study-comparing-the-gdcrw-and-the-ctcrw}{%
\subsection{Simulation study: Comparing the GDCRW and the
CTCRW}\label{simulation-study-comparing-the-gdcrw-and-the-ctcrw}}

Since the location process is constructed as an Euler--Maruyama discrete
time approximation to an underlying continuous time model, increasing
the number of latent states between observations should improve the
approximation. This was investigated by simulating from the underlying
continuous time model in the special case that reduces to the CTCRW. On
each simulated data set, the CTCRW was compared to the GDCRW with latent
states at the observations, and the GDCRW\(_i\) with additional latent
states evey \(i = 8, 4, 2\), and \(1\) time unit, respectively.

To compare the GDCRW with the CTCRW, 200 data sets were simulated from
the continuous time model defined by equations (\ref{eq:sdeV}) and
(\ref{eq:sdeX}) with \(\theta = 0\) and diagonal \(S\); that is, with
the CTCRW as the true model. To obtain locations, 250 time steps were
simulated from a mixture of an exponential and a normal distribution
(see Supplemental Material) to ensure both short and long time steps
between observations. Between the simulated time points, the velocity
and location processes were simulated using the Euler--Maruyama method
with 200 steps. For each of the 250 simulated locations, observations
were simulated from a bivariate normal with covariance \(0.1^2 I\). The
processes were simulated with \(\gamma_1=\gamma_2 = 0.9\), \(\mu = 0\),
and \(S_{11} = S_{22} = \exp(-2) \approx 0.135\). The models were fitted
to the data by maximum likelihood through the Laplace approximation
using the \texttt{R} package \texttt{TMB} \citep{kristensen2016a}. For
simplicity, the model was fitted with \(\gamma_1=\gamma_2\),
\(\theta = 0\), \(\mu = 0\), and \(S_{12} = S_{21} = 0\).

\hypertarget{simulation-study-comparing-the-gdcrw-and-the-dcrw}{%
\subsection{Simulation study: Comparing the GDCRW and the
DCRW}\label{simulation-study-comparing-the-gdcrw-and-the-dcrw}}

Above it was established that the DCRW was a special case of the GDCRW.
One of the ways the DCRW was generalized was that the GDCRW allowed
latent states at arbitrary time points. In contrast, the DCRW only
allowed equidistant time steps in discretizing the underlying SDE. In
this simulation study the effect of the choice of discretization on
parameter estimates was investigated.

Data sets were simulated from two behavioural scenarios using the same
procedure as for the previous simulation study. For each scenario, 200
data sets were simulated from the continuous time model determined by
equations (\ref{eq:sdeV}) and (\ref{eq:sdeX}). The first scenario
emulated tortuous movement simulated with
\(\phi = \frac{\pi}{3} \approx 1.05\), \(\mu = 0\), and
\(\gamma_1 = \gamma_2 = 0.6\). The movement was simulated with diagonal
\(S\) such that \(S_{11} = S_{22} = \exp(-2)\). Measurements were
simulated from a normal distribution with covariance
\(\Sigma_o = 0.1^2 I_2\). The second scenario mimicked a more persistent
movement. Movement was simulated with \(\phi = 0\) and \(\gamma = 0.7\).
Similar to the first scenario, the movement was simulated with
\(S_{11} = S_{22} = \exp(-2)\) and the observations with
\(\Sigma_o = 0.1^2 I_2\).

For each data set, a total of 10 models were fitted consisting of the
GDCRW and the DCRW with \(N = 250, 500, 750, 1000,\) and 1250 latent
locations. In the DCRW, the latent locations were included
equidistantly. In the GDCRW, 250 latent locations were aligned with the
observations. Remaining latent locations were included recursively in
the middle of the longest time step. The models were fitted to the data
by maximum likelihood through the Laplace approximation. For simplicity,
the DCRW models were fitted with \(S_{11} = S_{22}\), and
\(S_{12} = S_{21} = 0\). The GDCRW models were fitted with
\(\gamma_1=\gamma_2\), \(\mu = 0\), \(S_{11} = S_{22}\), and
\(S_{12} = S_{21} = 0\) to reduce to the same parameters as the DCRW
models.

\hypertarget{simulation-study-effect-of-measurement-error-on-choice-of-time-steps}{%
\subsection{Simulation study: Effect of measurement error on choice of
time
steps}\label{simulation-study-effect-of-measurement-error-on-choice-of-time-steps}}

In the final simulation study, the GDCRW and DCRW is compared once more.
While the previous comparison focused on parameter estimates, this
simulation study compares the ability to accurately obtain location
estimates for different ratios of process variability and measurement
variability. Again, 200 data sets with 250 observations were simulated
from two behavioural scenarios using the same procedure as for the two
previous simulation studies. The same movement parameters as in the
previous simulation study was used, and the GDCRW and DCRW with 250
latent locations were fitted in the same manner as before; however, the
observations were simulated with covariances
\(\exp(-6)^2I, \exp(-5.5)^2I, \ldots,\exp(1.5)^2I, \exp(2)^2I\),
respectively. For each location for each track the relative distance to
the true locations were calculated for the GDCRW compared to the DCRW.
The DCRW locations were interpolated linearly to get values at the time
of observations.

\hypertarget{case-study-ringed-seal}{%
\subsection{Case study: Ringed seal}\label{case-study-ringed-seal}}

To illustrate its practical applicability, the GDCRW model was fitted to
the subadult ringed seal \emph{Pusa hispida} data of
\citet{albertsen2015a} in this case study. The data set was chosen to
illustrate the movement model when substantial measurement errors are
present. The ringed seal data consists of \(N=3583\) locations collected
by the Argos satellite system, which is known to have substantial
non-Gaussian measurement errors. Three observations with location class
Z were excluded. In the first period of the track, the seal was moving
north-west, followed by two periods with restricted space use close to
land. Consequently, \citet{thygesen2017a} found indications that a
non-constant drift parameter was needed. To account for this behaviour,
the mean velocity parameters of the GDCRW were set to be location
dependent:\\
\[
\begin{aligned}
X_{t_i} =& X_{t_{i-1}} + \Delta_{t_i} \exp\left(-\left(\begin{array}{cc} -\log\gamma_1 & \theta \\ -\theta & -\log\gamma_2 \end{array}\right) \Delta_{t_{i-1}}\right) (X_{t_{i-1}} - X_{t_{i-2}})/\Delta_{t_{i-1}} \\
+& \Delta_{t_i}\left(I-\exp(-\Theta \Delta_{t_{i-1}})\right)\mu(X_{t_{i-1}}) + \Delta_{t_i} \epsilon_{t_i}
\end{aligned}
\] The bivariate mean velocity field
\(\mu(X_t) = (\mu_1(X_t), \mu_2(X_t))^T\) was modelled through a
\(7\times 7\) grid of random effects. Each grid point was related to two
random effects: one for the mean latitudinal velocity, \(\mu_1\), and
one for the mean longitudinal velocity, \(\mu_2\). Between the grid
points, the random effect values were interpolated by a search tree
approximation to local polynomial regression \citep{covafillr}. In this
case study, the number of random effects to construct the grids were
selected as a trade-off between computational cost and flexibility.
Locations of the movement process were estimated at every observation
and every three hours from the first observation, giving a total of 4764
location estimates each with two latent states. Following
\citet{albertsen2015a}, the measurements were modelled by a bivariate
t-distribution with scale matrices and degrees of freedom depending on
the location class of the observations. The model was fitted to the data
by maximum likelihood through the Laplace approximation using
\texttt{TMB}.

Having a location dependent drift in the model, the animal movement may
be related to available resources or hydrographic variables. The model
cannot be estimated using the DCRW, since the DCRW does not include a
drift parameter. Furthermore, the model cannot easily be estimated using
CTCRW without assuming constant movement between states, in which case
the CTCRW is a special case of the GDCRW (see \nameref{gdcrwVSctcrw}).

\hypertarget{results}{%
\section{Results}\label{results}}

\hypertarget{simulation-study-comparing-the-gdcrw-and-the-ctcrw-1}{%
\subsection{Simulation study: Comparing the GDCRW and the
CTCRW}\label{simulation-study-comparing-the-gdcrw-and-the-ctcrw-1}}

In the first simulation study, the GDCRW was compared to the CTCRW, when
the CTCRW was the true data generating model. The CTCRW models both the
velocity and location giving a total of 1000 latent variables: 4 for
each location. In contrast, the GDCRW only has 500 latent variables: 2
for each location. The mean of the time step distribution used in the
simulation study was approximately 1.99, giving an average length of a
simulated trajectory of 497.5 time units. Therefore, the GDCRW\(_8\)
would include 63 additional location estimates on average, corresponding
to a total of 313 latent variables. Likewise, the GDCRW\(_4\),
GDCRW\(_2\), and GDCRW\(_1\) would on average include 125, 249, and 498
additional locations, respectively, corresponding to 750, 998, and 1496
latent variables.

\begin{figure}
\centering
\includegraphics{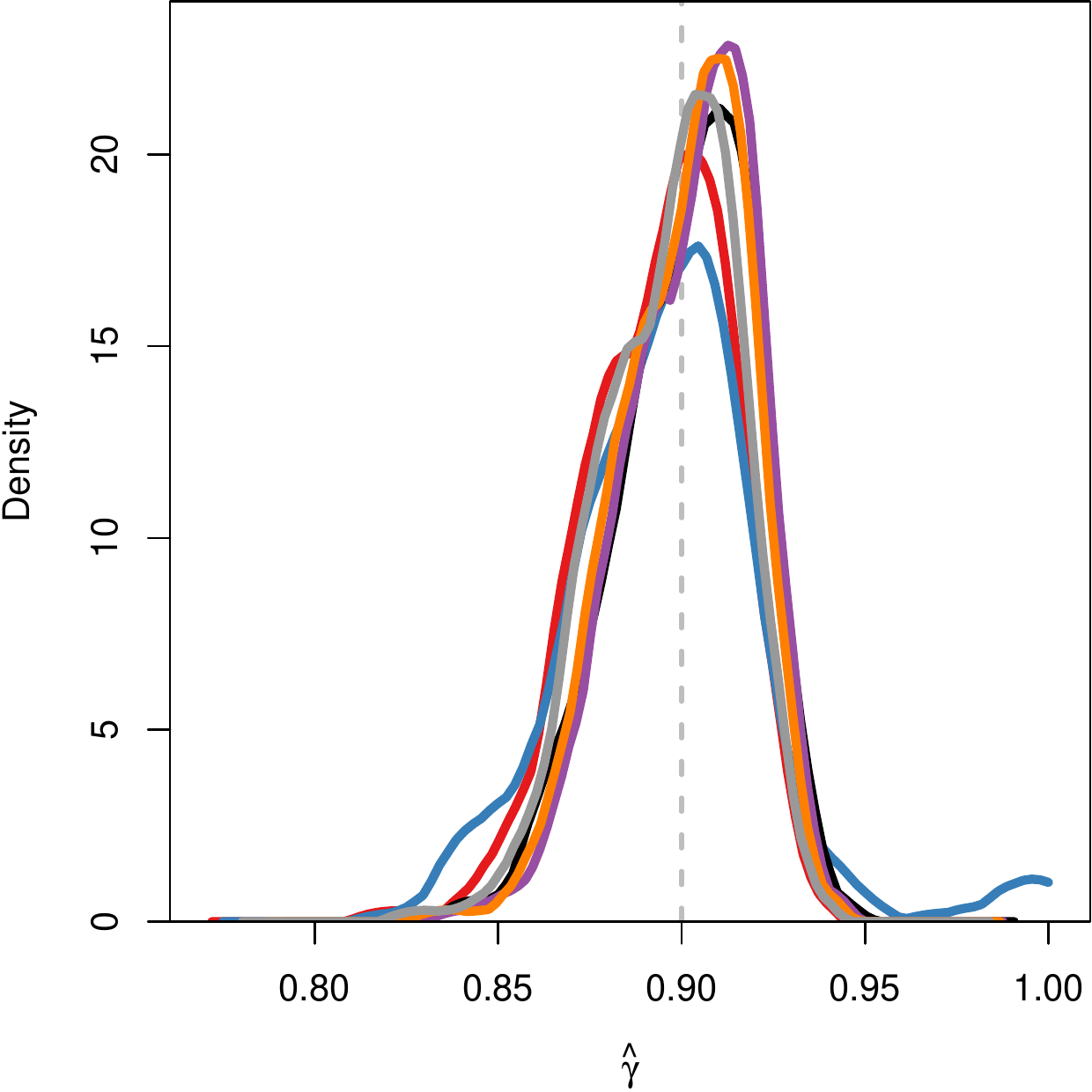}
\caption{Density of \(\gamma\) estimates in the simulation study for the
models: CTCRW (red line), GDCRW (blue line), GDCRW\(_8\) (black line),
GDCRW\(_4\) (purple line), GDCRW\(_2\) (orange line), and GDCRW\(_1\)
(grey line). Grey dashed line indicates the true parameter
value.\label{fig:simCTCRWest}}
\end{figure}

Regardless of the different number of latent variables, all six
estimation models provided \(\gamma\) estimates close to the true value
(\autoref{fig:simCTCRWest}). For the true continuous time model, the
CTCRW, the average of the 200 estimates was 0.894, whereas it was 0.897
for the discrete time approximation, the GDCRW, with latent states only
at the time of the observations. When additional latent states were
included every 8, 4, 2, and 1 time unit, the average of the estimated
\(\gamma\)s were 0.901, 0.903, 0.9, and 0.897, respectively. Although
all models gave estimates close to the true value, the standard
deviation of the estimates was higher for the GDCRW than for the five
other models. For the GDCRW the standard deviation of the 200 estimates
was 0.028, whereas it was 0.019 for the CTCRW, and 0.018, 0.017, 0.017,
and 0.018 for the GDCRW\(_8\), GDCRW\(_4\), GDCRW\(_2\), and
GDCRW\(_1\), respectively.

\begin{figure}
\centering
\includegraphics{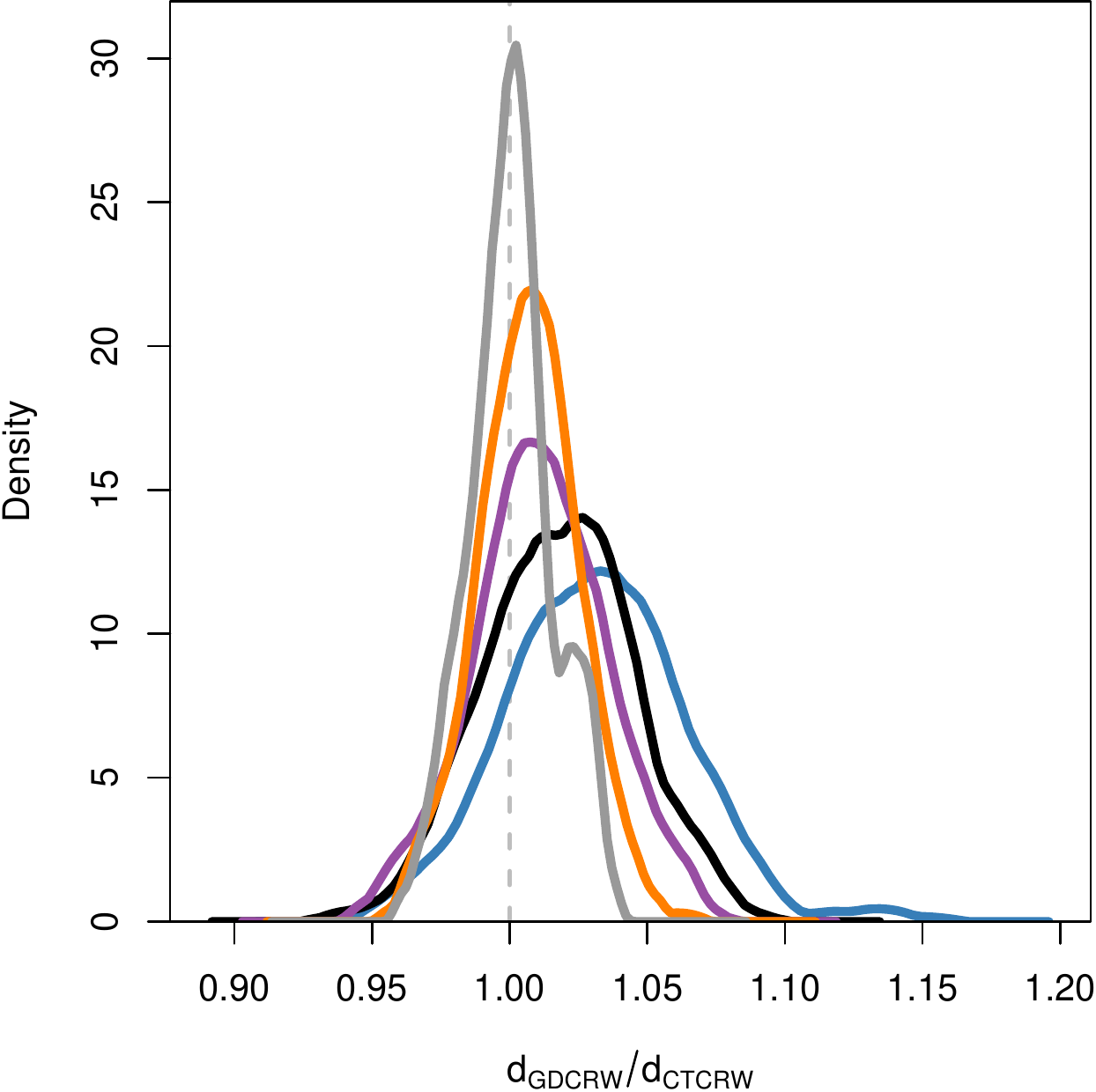}
\caption{Density of track-wise median distance between estimated
locations and true simulated locations in the simulation study for the
models: GDCRW (blue line), GDCRW\(_8\) (black line), GDCRW\(_4\) (purple
line), GDCRW\(_2\) (orange line), and GDCRW\(_1\) (grey line) relative
to the CTCRW. A value of one (grey dashed line) indicates the median
distance is the same as for the CTCRW.\label{fig:simCTCRWdist}}
\end{figure}

Besides the ability to re-obtain the true parameters, the models were
compared on the distance between the estimated locations and the true
simulated locations \autoref{fig:simCTCRWdist}. To compare the models,
this distance was calculated for each location for each track for the
GDCRW, GDCRW\(_8\), GDCRW\(_4\), GDCRW\(_2\), and GDCRW\(_1\), and
divided by the distance obtained for the CTCRW. Finally, the median
relative distance was obtained. Whereas the models performed similarly
in estimating \(\gamma\), the CTCRW performed better than the other
models in obtaining precise location estimates; however, the GDCRW
models improved when additional latent states were included, as
expected. For the GDCRW, 16\% of the simulations had a relative distance
compared to the CTCRW that was less than one; that is, where at least
half of the estimated locations were closer to the true location than
for the estimated locations from the CTCRW. For the GDCRW\(_8\), 24\% of
the simulations had a relative distance less than one, while it was 31,
36, and 47\%, respectively, for the GDCRW\(_4\), GDCRW\(_2\), and
GDCRW\(_1\). Further, the range of median relative distances narrowed
when the number of latent states increased. When no additional latent
states were included, the length of the range was 0.188 (0.958;1.146).
Based on these 200 simulations, the range gradually narrowed for the
GDCRW\(_8\) (0.143; min: 0.958; max: 1.146), GDCRW\(_4\) (0.116; min:
0.953; max: 1.069), GDCRW\(_1\) (0.099; min: 0.962; max: 1.061), and
GDCRW\(_1\) (0.072; min: 0.964; max: 1.036).

\hypertarget{simulation-study-comparing-the-gdcrw-and-the-dcrw-1}{%
\subsection{Simulation study: Comparing the GDCRW and the
DCRW}\label{simulation-study-comparing-the-gdcrw-and-the-dcrw-1}}

\begin{figure}
\centering
\includegraphics{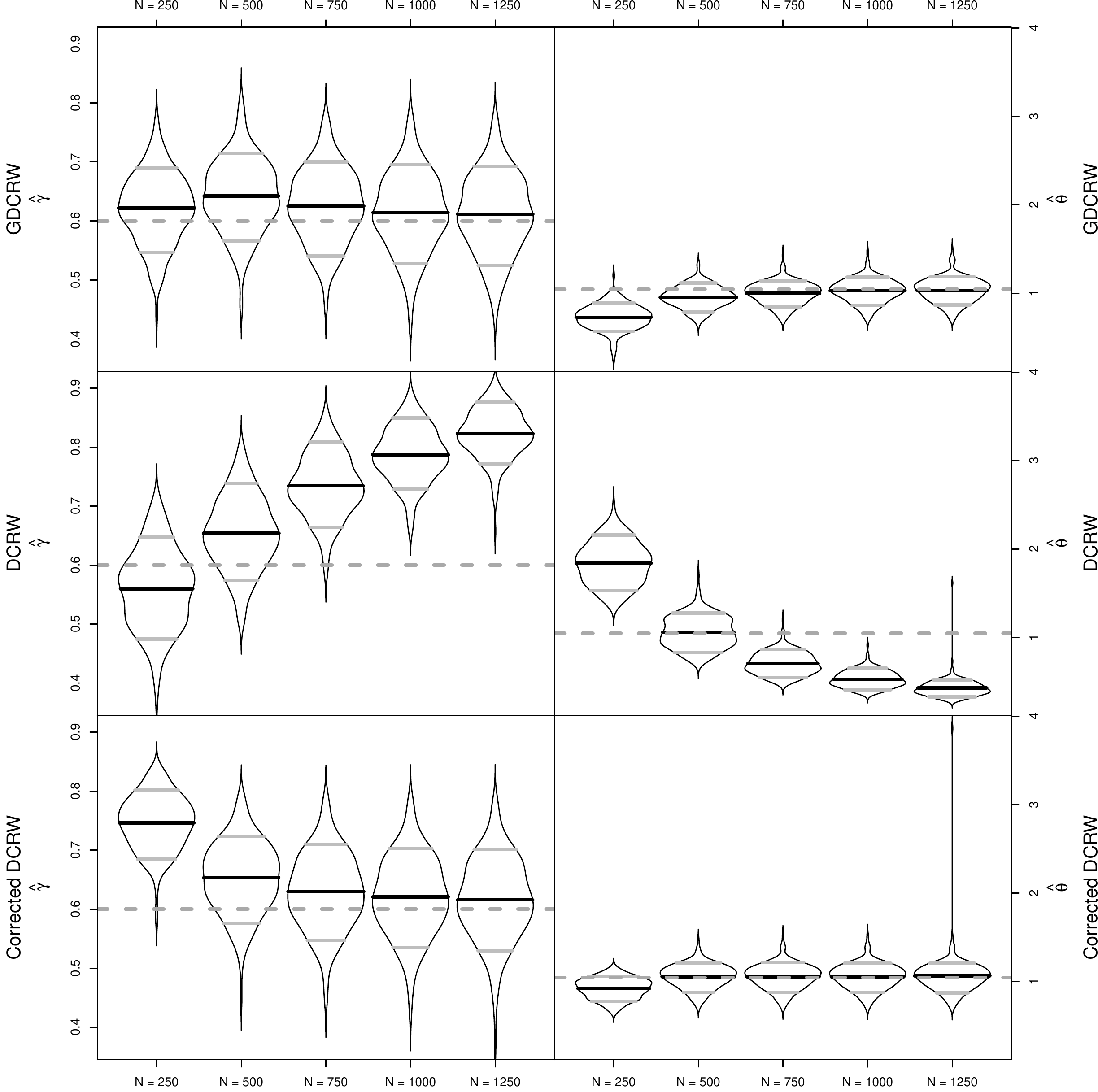}
\caption{Violin plots of \(\gamma\) and \(\theta\) estimates in the
tortuous movement simulations for the GDCRW, DCRW, and the DCRW
corrected by equation (\ref{eq:reggdcrw}) with
\(N = 250, 500, 750, 1000, 1250\) latent states and 250 observations.
Grey dashed lines indicate true parameter values. In the violin plot,
black lines indicate the mean of estimates and grey lines indicate 10
and 90 \% quantiles of estimates.\label{fig:simDCRWestT}}
\end{figure}

In the second simulation study, the parameter estimates from the GDCRW
and DCRW were compared in two movement scenarios. In the tortuous
movement scenario, the GDCRW provided \(\gamma\) estimates close to the
true value for all five number of latent locations with a slight
tendency to overestimate the value (\autoref{fig:simDCRWestT}). The
average \(\gamma\) estimates (standard deviation) were 0.622 (0.059),
0.643 (0.061), 0.625 (0.062), 0.614 (0.066), and 0.611 (0.067) for the
GDCRW with \(N = 250, 500 ,750, 1000,\) and 1250 latent locations,
respectively. A difference was, however, seen in for the \(\theta\)
estimates. Their accuracy increased with the number of latent locations.
The average estimates were 0.729 (0.139), 0.956 (0.124), 0.999 (0.124),
1.028 (0.129), and 1.033 (0.134), respectively. Unlike for the GDCRW,
the estimates obtained directly from the DCRW were not close to the true
values. For these models, the average \(\gamma\) estimates were 0.56
(0.067), 0.654 (0.061), 0.734 (0.053), 0.787 (0.046), and 0.823 (0.041)
with \(N = 250, 500 ,750, 1000,\) and 1250 latent locations,
respectively, while the average \(\theta\) estimates were 1.841 (0.237),
1.059 (0.178), 0.705 (0.124), 0.528 (0.094), and 0.427 (0.113).
Correcting the estimates obtained from the DCRW by equation
(\ref{eq:reggdcrw}), the accuracy of the estimates increased with the
number of latent locations. Then, the average \(\gamma\) estimates were
0.746 (0.046), 0.653 (0.059), 0.63 (0.064), 0.621 (0.066), and 0.616
(0.069), respectively, while the average \(\theta\) estimates were 0.921
(0.114), 1.057 (0.133), 1.055 (0.136), 1.054 (0.138), and 1.066 (0.241).

\begin{figure}
\centering
\includegraphics{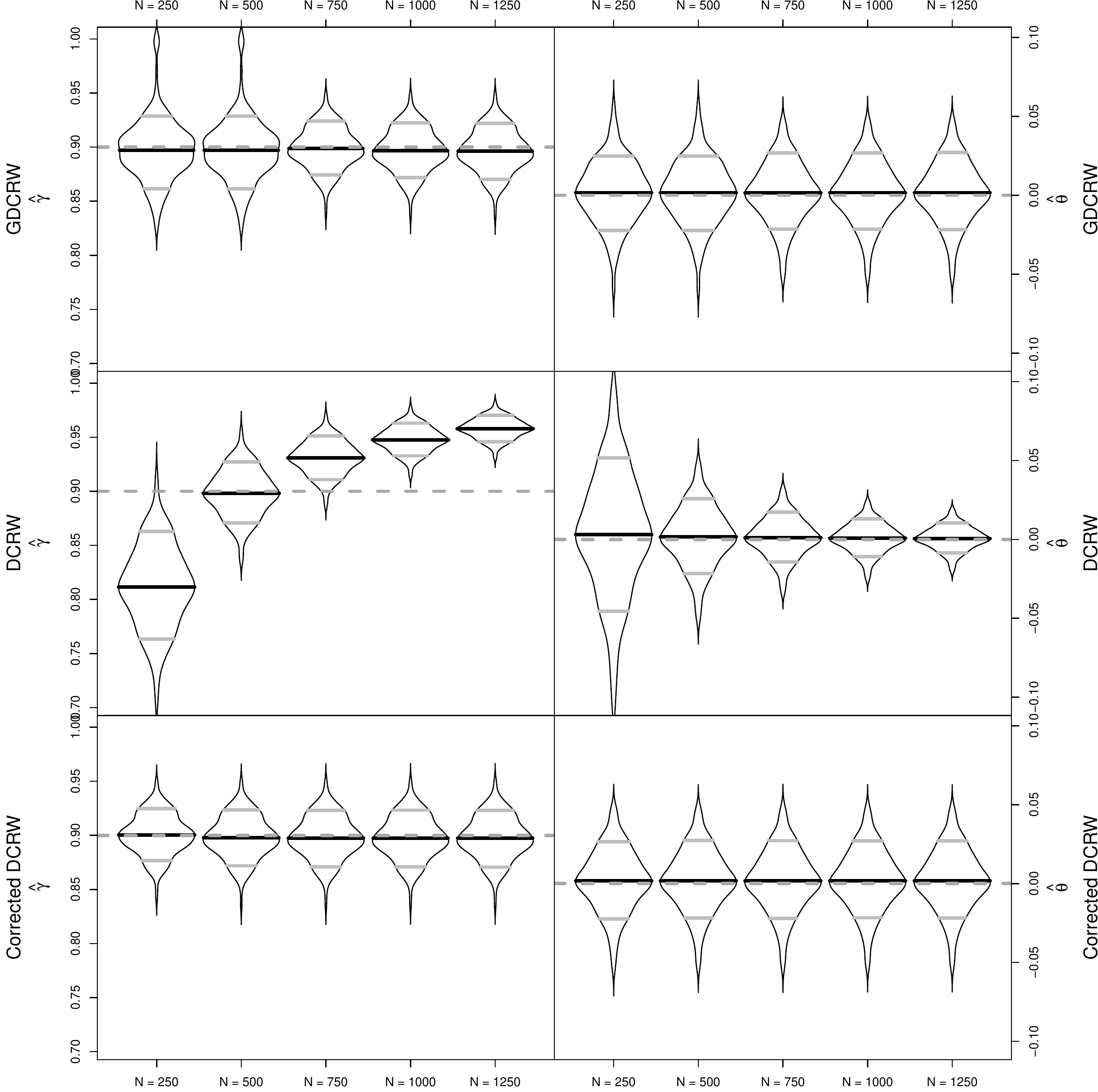}
\caption{Violin plots of \(\gamma\) and \(\theta\) estimates in the
persistent movement simulations for the GDCRW, DCRW, and the DCRW
corrected by equation (\ref{eq:reggdcrw}) with
\(N = 250, 500, 750, 1000, 1250\) latent states and 250 observations.
Grey dashed lines indicate true parameter values. In the violin plot,
black lines indicate the mean of estimates and grey lines indicate 10
and 90 \% quantiles of estimates.\label{fig:simDCRWestD}}
\end{figure}

In the persistent movement scenario, the GDCRW and the DCRW corrected by
equation (\ref{eq:reggdcrw}) provided estimates close to the true values
for both parameters for all five number of latent locations
(\autoref{fig:simDCRWestD}). For the GDCRW, the average \(\gamma\)
estimates were 0.897 (0.028), 0.897 (0.028), 0.899 (0.019), 0.897
(0.019), and 0.896 (0.019), respectively, and the average \(\theta\)
estimates were 0.002 (0.02), 0.002 (0.02), 0.002 (0.019), 0.002 (0.019),
and 0.002 (0.019). Likewise, the average DCRW estimates corrected by
equation (\ref{eq:reggdcrw}) were 0.9 (0.019), 0.898 (0.019), 0.898
(0.019), 0.897 (0.019), and 0.897 (0.019), respectively, for \(\gamma\),
and 0.002 (0.019), 0.002 (0.019), 0.002 (0.019), 0.002 (0.019), and
0.002 (0.019), respectively, for \(\theta\). Similarly to the tortuous
movement scenario, the estimates obtained directly from the DCRW were
not close to the true values. For the DCRW models, the average
\(\gamma\) estimates were 0.812 (0.037), 0.898 (0.022), 0.931 (0.015),
0.948 (0.011), and 0.958 (0.009) with \(N = 250, 500 ,750, 1000,\) and
1250 latent locations, respectively, while the average \(\theta\)
estimates were 0.003 (0.038), 0.002 (0.019), 0.001 (0.012), 0.001
(0.009), and 0.001 (0.007).

\hypertarget{simulation-study-effect-of-measurement-error-on-choice-of-time-steps-1}{%
\subsection{Simulation study: Effect of measurement error on choice of
time
steps}\label{simulation-study-effect-of-measurement-error-on-choice-of-time-steps-1}}

\begin{figure}
\centering
\includegraphics{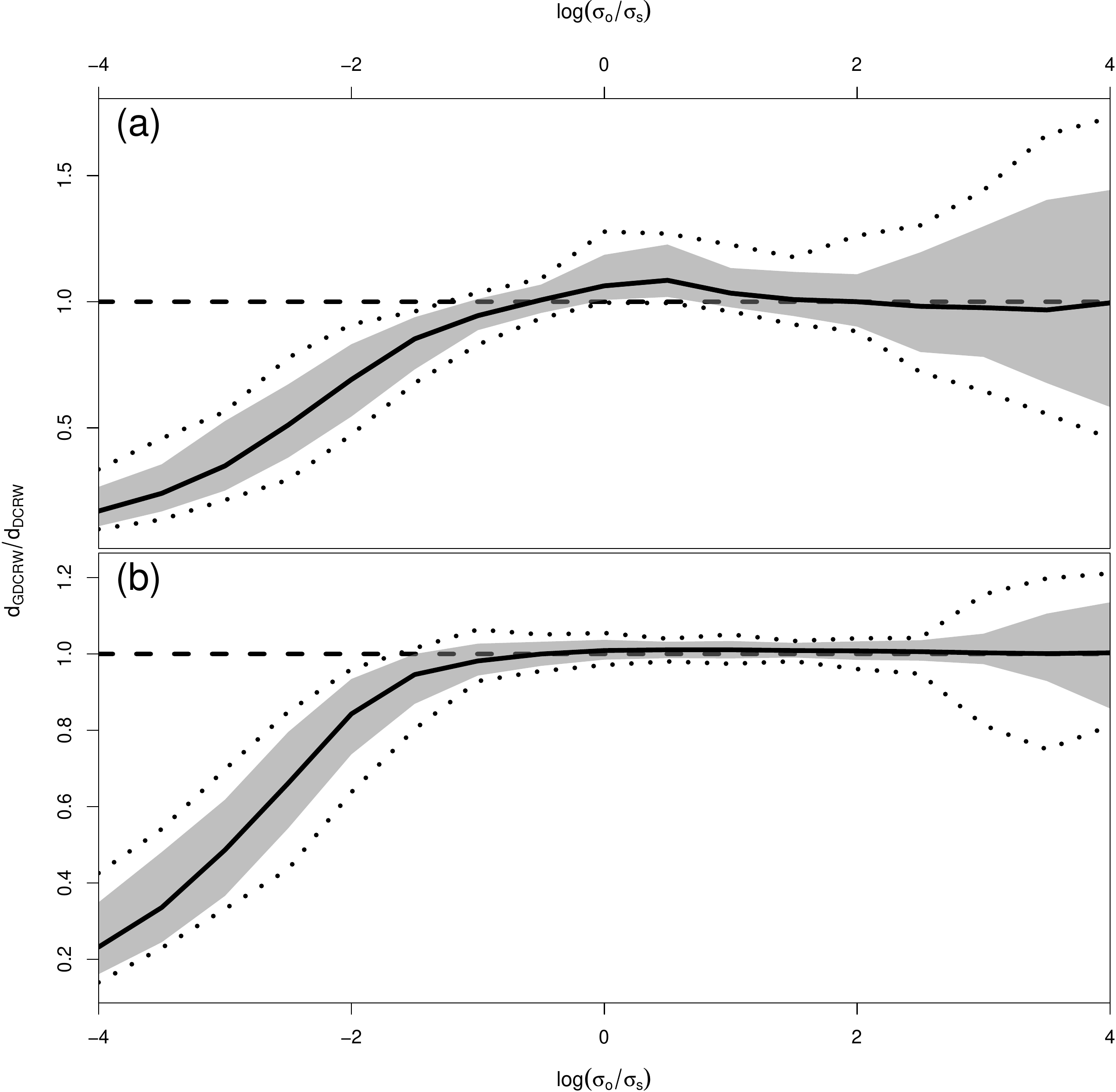}
\caption{Minimum (lower dotted black line), maximum (upper dotted black
line), median (full black line), and 95\% range (grey area) of
track-wise median distances between estimated locations and true
simulated locations for the GDCRW relative to the DCRW in the tortuous
(a) and persistent (b) simulated movement scenarios as a function of the
ratio between observational standard deviation (\(\sigma_o\)) and
movement standard deviation (\(\sigma_s\)). Values below 1 (black dashed
line) indicates that at least half of the estimated location in the
GDCRW were closer to the true value than for the
DCRW.\label{fig:simCompDist}}
\end{figure}

In the third simulation study, the GDCRW and DCRW were compared again.
This simulation study compared location estimates for different levels
of measurement error. When the measurement standard deviation was low,
the GDCRW performed notably better than the DCRW
(\autoref{fig:simCompDist}). In both scenarios, all 200 simulations had
a median relative distance less than 1; that is, at least half of the
locations were better estimated by the GDCRW than the DCRW. As the
measurement standard deviation increases relative to the movement
standard deviation, the median relative distance tends to one,
suggesting that the GDCRW and the DCRW estimates similarly. When the
measurement and movement standard deviation were equal, the DCRW
performed slightly better than the GDCRW.

\hypertarget{case-study-ringed-seal-1}{%
\subsection{Case study: Ringed seal}\label{case-study-ringed-seal-1}}

\begin{figure}
\centering
\includegraphics{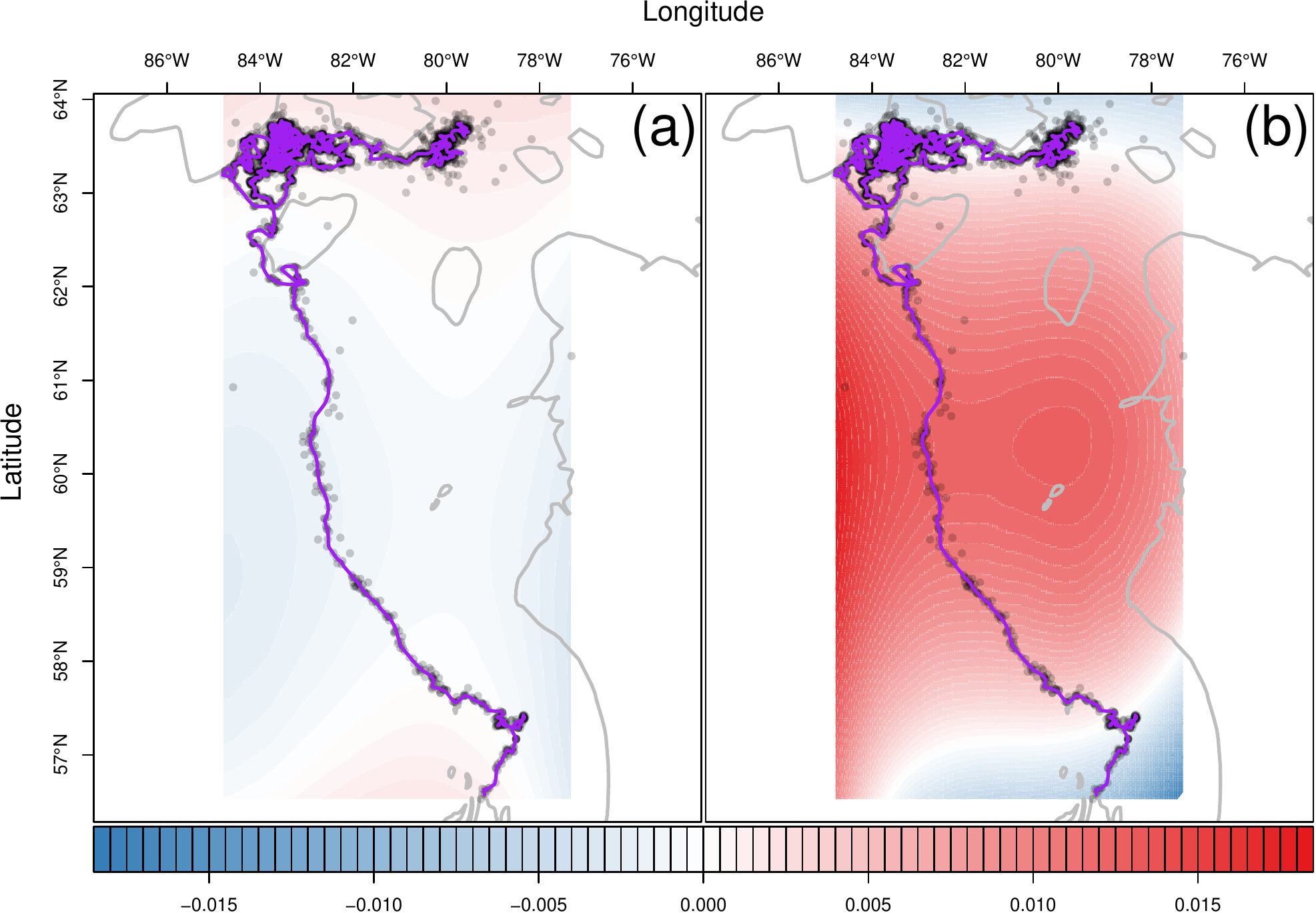}
\caption{Mean longitudinal (a) and latitudinal (b) velocity field of the
subadult ringed seal track with land (outlined by grey lines), Argos
satellite observations (dark grey dots), and fitted most probable track
given all observations (purple line). In the longitudinal field, red
values indicate an eastward attraction, while blue values indicate a
westward attraction. In the latitudinal field, red values indicate a
northward attraction, while blue values indicate a southward
attraction.\label{fig:seal}}
\end{figure}

In the case study, the GDCRW was fitted to a real data set collected by
the Argos system. The GDCRW was modified to include location dependent
drift and turning angle. The fitted model resulted in small estimated
longitudinal mean velocity field values (\autoref{fig:seal}). On the
contrary, the latitudinal mean velocity field was estimated to have a
southward drift when the seal was south of the 57.5\(^\circ\)N parallel.
Despite the southward drift, the seal then entered a large area of
northward drift between 57.5\(^\circ\)N and 63.5\(^\circ\)N. North of
the 63.5\(^\circ\)N parallel, the latitudinal mean velocity field was
estimated to have a southward attraction, keeping the seal close to the
63.5\(^\circ\)N parallel. Further the model estimated an overall mean
turning angle, \(\theta\). This parameter was estimated to be close to
zero (\autoref{tab:est_seal}), whereas both the latitudinal and
longitudinal autocorrelation parameters were estimated to be close to
0.6.

\begin{table}

\caption{\label{tab:unnamed-chunk-16}Estimated movement parameters in the ringed seal case study.\label{tab:est_seal}}
\centering
\begin{tabular}[h]{lrr}
\toprule

  & Estimate & Standard Error\\
\midrule
$\gamma_{lat}$ & 0.608 & 0.027\\
$\gamma_{lon}$ & 0.601 & 0.028\\
$\theta$ & 0.027 & 0.019\\
$\Sigma_{lat,lon}/\sqrt{\Sigma_{lat,lat}\Sigma_{lon,lon}}$ & -0.018 & 0.052\\
$\log\Sigma_{lat,lat}$ & -9.189 & 0.114\\
$\log\Sigma_{lon,lon}$ & -7.486 & 0.114\\
\bottomrule
\end{tabular}
\end{table}

\hypertarget{discussion}{%
\section{Discussion}\label{discussion}}

The GDCRW introduced here generalizes the DCRW movement model of
\citet{jonsen2005a} in three ways. Firstly, whereas the DCRW handles
irregular observations by linear interpolation in the observational
model, the GDCRW allows modelling irregular observations directly in the
movement process. Through this construction, time-scale corrections of
the DCRW parameters were found by considering a time regular GDCRW.
Secondly, the GDCRW generalizes the DCRW by allowing different
auto-correlation parameters in the two coordinates. Hence, when the
rotation parameter is zero, the GDCRW resembles the integrated velocity
model of the CTCRW of \citet{johnson2008a}. Like the CTCRW and DCRW, the
GDCRW includes the random walk as a limiting case. When the
auto-correlation parameter approaches zero, the movement model
approaches that of a random walk, which has been used in several
studies, in particular when the data includes only daily observations
\citep[e.g.][]{nielsen2006a, lam2010a}. Thirdly, the GDCRW extends the
DCRW by including a drift term.

Including a drift term in the model is useful for animals moving
persistently between areas, such as the ringed seal analysed in the case
study. In the case study, the drift term was modelled by a location
dependent field. Having location or time dependent parameters in the SDE
greatly complicates the calculations to obtain an analytical solution
and in turn a fully continuous time model. Therefore, location dependent
mean velocities could not easily be implemented in the CTCRW which has
previously been used to analyse the track
\citep{albertsen2015a, thygesen2017a}. Nonetheless, using the discrete
time model, or a discrete time approximation, it is straightforward to
extend the movement model with a location dependent drift. Using the
same methods, spatial covariates such as sea surface temperature or
depth could be included.

While the mean turning angle was estimated to be close to zero using the
entire track, this may not be the case if, for instance, a time varying
parameter or behavioural switching was included. Changes in behavioural
states, \(S_{t_i} \in \{1,2,\ldots,n\}\), can be modelled by a Markov
Chain such that \[
P(S_{t_i} = k \mid S_{t_{i-1}} = j) = (\exp(Q\Delta_{t_i}))_{jk}
\] where \(Q\) is an \(n \times n\) matrix such that \(Q_{jk} > 0\) for
\(j\neq k\) and \(Q_{jj} = -\sum_{\{k:j\neq k\}} Q_{jk}\). The movement
model is modified to have different parameters depending on the current
behavioural state \[
X_{t_i} = X_{t_{i-1}} + \Delta_{t_i} \exp(-\Theta_{S_{t_i}} \Delta_{t_i}) (X_{t_{i-1}} - X_{t_{i-2}})/\Delta_{t_{i-1}} + \Delta_{t_i}\left(I-\exp(-\Theta_{S_{t_i}} \Delta_{t_i})\right)\mu_{S_{t_i}} + \Delta_{t_i} \epsilon_{t_i}.
\] This model generalizes the behavioural switching movement models of
\citet{jonsen2005a} and \citet{whoriskey2017a}. It assumes that
switching only occurs at the predetermined time points; however, as it
has been shown in both the case study and simulation studies, latent
locations can be included either at regular time points, or at the time
of observations and any time between them. While the extension to
include behavioural switching is simple, maximum likelihood estimation
is greatly complicated when measurement error is present.

The GDCRW can be constructed as a discretization of a generalization of
the CTCRW. The first simulation study comparing the GDCRW and CTCRW
suggests that using the analytical continuous time solution, when it can
be found, generally provides better results than a discrete time
approximation. Nonetheless, the discrete time approximation provided
accurate parameter estimates, even when no additional latent states were
included to improve the approximation. While the GDCRW was outperformed
by the CTCRW in estimating true locations when the CTCRW was the true
model, the first simulation study illustrated that the performance of
the GDCRW could be improved to closely resemble the CTCRW by including
additional latent states between observations.

The second simulation study compared the DCRW and the GDCRW in their
ability to re-estimate the movement parameters of the underlying
continuous time model. In both the tortuous and persistent movement
scenarios, the GDCRW performed well; however, in the tortuous movement
scenario, the performance was notably improved by adding additional
latent locations between observations. Further, it became evident that
the parameter estimates of the DCRW are highly dependent on the selected
time steps between latent locations. Nevertheless, the estimates could
be corrected by equation (\ref{eq:reggdcrw}): the GDCRW with regular
time steps. This correction provided estimates close to the true values
in all cases except in the tortuous movement case with 250 and 500
latent locations. Clearly, the ability to have time scale independent
movement parameters is a key feature of the GDCRW, regardless of whether
it is used with regular or irregular time steps. Time scale independent
movement parameters allow analysis of several animals without using the
same time steps, and further allow comparing results from previous
studies using the DCRW even if different time steps are used.

Besides accuracy of the estimates, the DCRW and GDCRW were compared on
their ability to reconstruct the movement tracks. The third simulation
study showed that modelling time-irregular data through a time-irregular
movement could increase the accuracy of predicted locations. This is
consistent with previous findings comparing the DCRW and CTCRW, which
showed that the continuous time model performed better if both movement
models were combined with t-distributed measurement errors
\citep{albertsen2015a}. Overall, using the GDCRW with irregular time
steps performed well in both scenarios compared to the DCRW. For small
measurement standard deviations, the irregular time steps outperformed
the regular time steps with interpolation. For medium measurement
standard deviations, the two approaches performed equally in the
persistent movement scenario, whereas the DCRW slightly outperformed the
GDCRW in the tortuous movement case. For large measurement standard
deviations, the two approaches performed equally in both scenarios.
Combined with the previous simulation study, these results suggest that
having latent locations at the time of observations generally performs
well compared to the DCRW with equidistant time steps in the movement;
keeping in mind that both improves as approximations to the same
underlying continuous time movement model as the number of latent
locations increases.

Directly modelling irregular data eases interpretation and comparison of
movement parameters because they are scaled by the time steps. From
equation (\ref{eq:reggdcrw}), parameters of the DCRW and GDCRW can be
transformed to any time scale. However, the simulation study shows that
fitting the DCRW with a poorly chosen time step introduces bias. This
poses a problem in population and meta studies. With a discrete time
model, all animals must be fitted with the same time steps to compare
movement parameters. However, if the same time step is not optimal for
all animals, bias can be introduced. Using an irregular or continuous
time model scales the parameters to a common time scale, even when the
animals are observed, or behave, at different time scales.

\bibliographystyle{apalike-doi.bst}
\bibliography{manuscript.bib}

\newpage

\hypertarget{appendix-s1-solving-the-sde}{%
\section{Appendix S1: Solving the
SDE}\label{appendix-s1-solving-the-sde}}

To solve the SDE, \[
\text{d}V_t = - \Theta (V_t-\mu) \text{d}t + S dB_t,
\] consider the process \[
Y_t = e^{\Theta t}(V_t-\mu)
\] It\(\^o\)'s formula yields \[
\begin{aligned}
dY_t &= \Theta e^{\Theta t} V_t d_t + e^{\Theta t}dV_t \\
&= \Theta e^{\Theta t} V_t d_t + e^{\Theta t} \left( - \Theta (V_t-\mu) d_t + S dB_t \right) \\
&= e^{\Theta t} S dB_t
\end{aligned}
\] Hence, \[
e^{\Theta t}(V_{t}-\mu) = (V_0-\mu) + \int_0^{t} e^{\Theta u} S dB_u
\] which implies \[
\begin{aligned}
V_{t}-\mu &= e^{-\Theta t}(V_0-\mu) + e^{-\Theta t} \int_0^{t} e^{\Theta s} S dB_s \\
&= e^{-\Theta t}V_0 + \left(I-e^{-\Theta t}\right)\mu + \int_0^{t} e^{-\Theta (t-s)} S dB_s
\end{aligned}
\] Since this is a sum of a deterministic term and an integral of a
deterministic function with respect to a Wiener process with Gaussian
increments, the distribution is Gaussian. The mean of the increment is
\[
\begin{aligned}
E(V_{t} \mid V_{0}) &= E\left(e^{-\Theta t}V_0 + \left(I-e^{-\Theta t}\right)\mu \mid V_{0} \right) + E\left( \int_0^{t} e^{-\Theta (t-s)} S dB_s \mid V_{0}\right) \\
&= e^{-\Theta t} V_0 + \left(I-e^{-\Theta t}\right)\mu \\
\end{aligned}
\]

Using It\(\^o\) isometry, the variance is \[
\begin{aligned}
Var\left( V_{t} \mid V_{0} \right) &= Var\left(e^{-\Theta t}V_0  + \left(I-e^{-\Theta t}\right)\mu + \int_0^{t} e^{-\Theta (t-s)}SdB_s  \mid V_{0} \right) \\
&= Var\left(\int_0^{t} e^{-\Theta (t-s)} S dB_s \mid V_{0} \right) \\
&= Var\left(\int_0^{t} e^{-\Theta (t-s)} S dB_s \right) \\
&= E\left(\left(\int_0^{t} e^{-\Theta (t-s)} S dB_s\right)^2 \right) - E\left(\int_0^{t} e^{-\Theta (t-s)} S dB_s \right)^2\\
&= E\left(\left(\int_0^{t} e^{-\Theta (t-s)} S dB_s\right)^2 \right) \\
&= \int_0^t e^{-\Theta (t-s)} SS^T \left(e^{-\Theta (t-s)}\right)^T ds \\
&= \int_0^t e^{-\Theta (t-s)} \Sigma \left(e^{-\Theta (t-s)}\right)^T ds
\end{aligned}
\] Now \citep{gentle2007a}, \[
\begin{aligned}
\operatorname{vec}(Var\left( V_{t} \mid V_{0} \right)) &= \int_0^t e^{-\Theta (t-s)} \otimes e^{-\Theta (t-s)} \operatorname{vec}(\Sigma) ds \\
&= \int_0^t e^{-\Theta\oplus\Theta (t-s)} ds \operatorname{vec}(\Sigma) \\
&= (\Theta\oplus\Theta)^{-1} \left(I - e^{-\Theta\oplus\Theta t}\right) \operatorname{vec}(\Sigma)
\end{aligned}
\] where \(\oplus\) denotes the Kronecker sum,
\(A \oplus B = A \otimes I_B + I_A \otimes B\).

Defining the matrix \(C\) such that
\(\operatorname{vec}(C) = (\Theta\oplus\Theta)^{-1} \operatorname{vec}(\Sigma)\),

\[
\begin{aligned}
\operatorname{vec}(Var\left( V_{t} \mid V_{0} \right)) &= (\Theta\oplus\Theta)^{-1} \operatorname{vec}(\Sigma) - (\Theta\oplus\Theta)^{-1} e^{-\Theta\oplus\Theta t} \operatorname{vec}(\Sigma) \\
&= (\Theta\oplus\Theta)^{-1} \operatorname{vec}(\Sigma) -e^{-\Theta\oplus\Theta t}  (\Theta\oplus\Theta)^{-1} \operatorname{vec}(\Sigma) \\
&= \operatorname{vec}(C) - e^{-\Theta\oplus\Theta t} \operatorname{vec}(C) \\
&= \operatorname{vec}(C) - e^{-\Theta t}\otimes e^{-\Theta t} \operatorname{vec}(C)
\end{aligned}
\]

Hence,

\[
Var(V_t \mid V_0) = C - e^{-\Theta t} C e^{-\Theta^T t}
\]

\hypertarget{appendix-s2-simulation-method-details}{%
\section{Appendix S2: Simulation method
details}\label{appendix-s2-simulation-method-details}}

\begin{figure}
\centering
\includegraphics{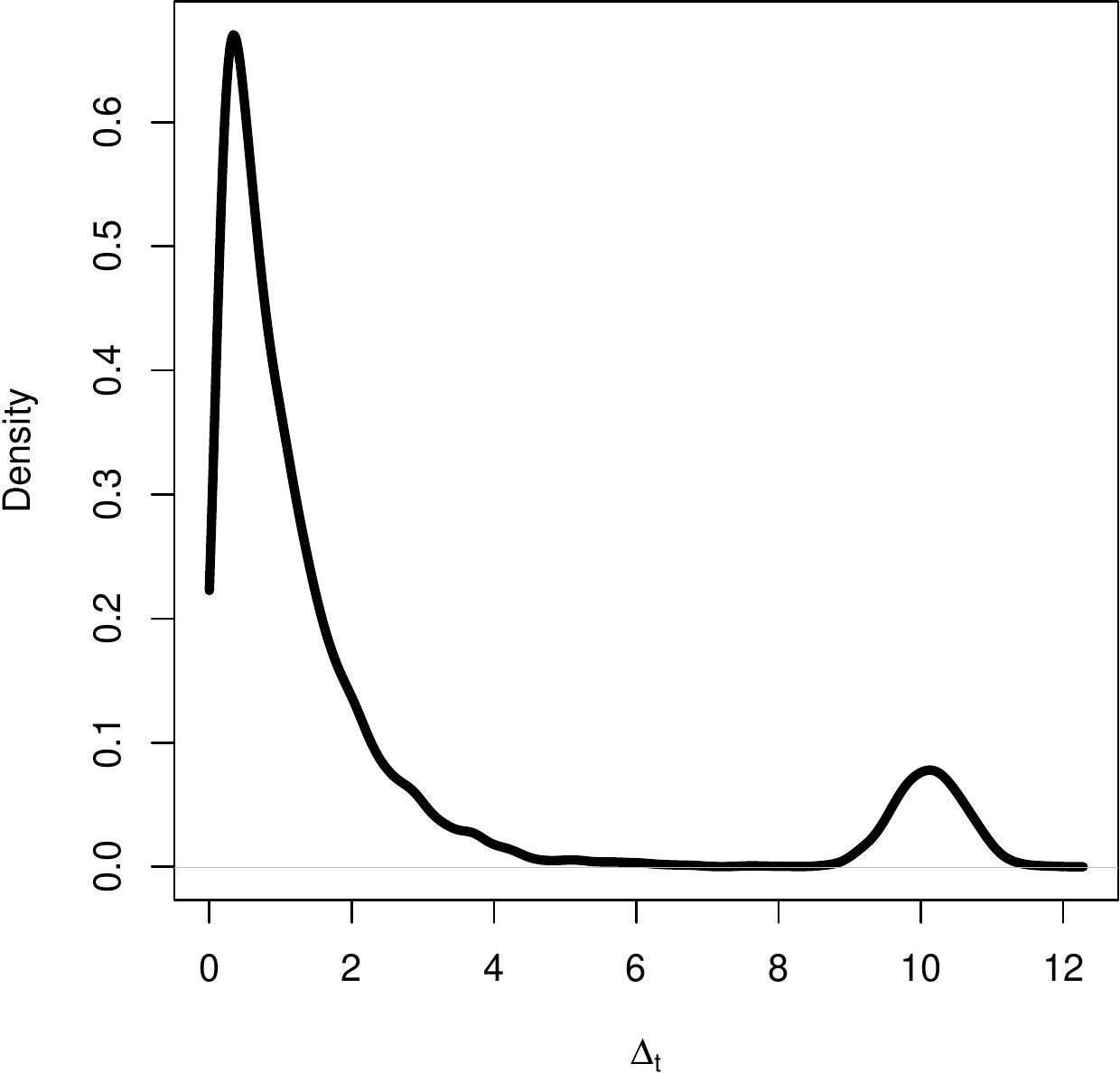}
\caption{Probability density function used to simulate time
steps.\label{fig:dt}}
\end{figure}

To simulate from the process, \[
\begin{aligned}
\text{d}V_t &= - \Theta (V_t-\mu) \text{d}t + S dB_t
\text{d}X_t &= V_t\text{d}t 
\end{aligned}
\] time steps,\(\Delta_{t_i} = t_i - t_{i-1}\), between observations
where simulated from a mixture of an exponential distribution and a
normal distribution. To simulate from the mixture, a uniform random
variable, \(U\sim unif(0,1)\), an exponentially distributed random
variable, \(\delta_0\sim exp(1)\), and a Gaussian random variable,
\(\delta_1\sim N(10,0.5^2)\), were simulated. Then the time step was \[
\Delta_{t_i} = \left\{\begin{array}{cc} 0.1 + \delta_0 & U < 0.9 \\ 0.1 + \max(\delta_1,0.01) & U \geq 0.9 \end{array}\right.
\] The resulting density function is seen in \autoref{fig:dt}.

\begin{figure}
\centering
\includegraphics{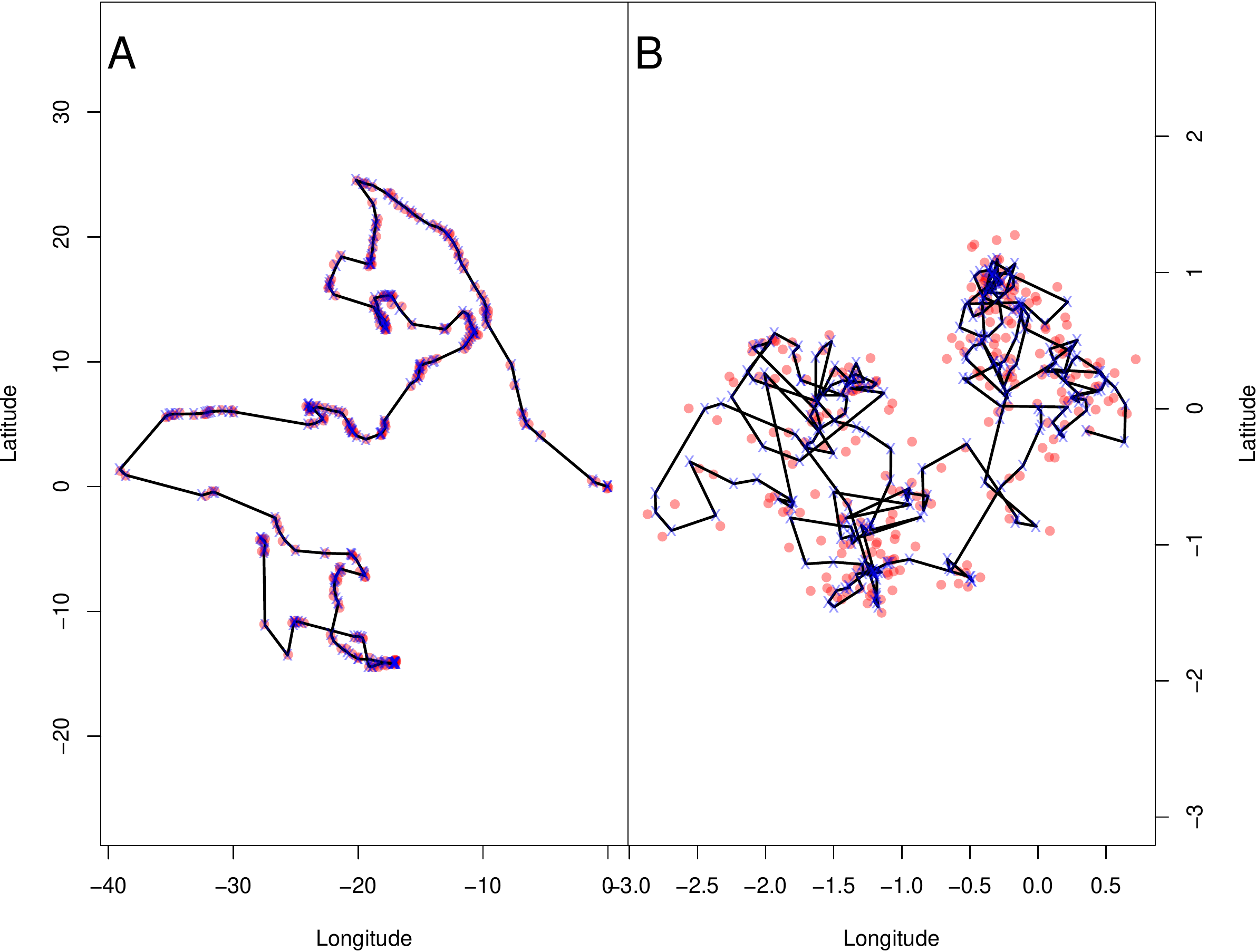}
\caption{Simulated trajectories (blue crosses connected by black lines)
and observations (red points) from the SDE model in the two scenarios:
persistent (A) and tortuous (B) movement.\label{fig:simtraj}}
\end{figure}

Between two time points \(t_i\) and \(t_{i+1}\), the processes were
simulated using the Euler--Maruyama approximation, \[
\begin{aligned}
V_{s(j+1)} &= V_{s(j)} - \Theta (V_n-\mu) \Delta_{{s(j+1)}} + S \eta_{{s(j+1)}}
X_{s(j+1)} &= X_{s(j)} + V_{s(j+1)} \Delta_{{s(j+1)}}
\end{aligned}
\] with \(s(j) = t_i + (n)/199 \cdot \Delta_{t_{i+1}}\),
\(j = 0,1,\ldots 199\). Simulated examples can be seen in
\autoref{fig:simtraj}

\end{spacing}
\end{document}